# Use of computer by secondary school students

Dr.kirti matliwala

Assistant professor,

Education department (SF),

VNSGU,

Surat

**Abstract**

This paper reports on the outcomes of a survey implemented in secondary schools. The survey identified the types of access and use of computers by students. It was found that the students had significant access to computers but they were not skilled in many features of computer use. Computers were used for a range of activities, some educational and others recreational by some students. Gender differences in computer use were not seen. The study highlights the changing scenario about uses of computer technology by students.

 **Introduction**

The lives of teens have undergone significant changes in the past few years by the acceptance of computers. We know that the use of computers by teens and their skills affect their life. Teens have extensive exposure to computers in their school and out-of-school contexts and their concomitant dispositions to learning and activity. Nowadays digital technology has been an integral part of their lives. Characteristics of these students are quite different from previous generations because of the social and technological conditions within which they are developing. There is an international recognition of the potential of computer technology to create new learning and environments (Cuban, 2003).

Prensky (2001) has been a notable writer on the phenomenon of the digital native. He argued that this generation, having grown up immersed in technology, has begun to think differently from other generations (Prensky, 2005). Their homes have computer technology in all facets of gadgetry—the remote control for the television, the programmable microwave, mobile telephone, computers and digital games. Prensky (2005) argues that digital natives are more connected than other generations through technologies such as mobile phones, email and chat lines. Communication is a much more connected and global experience for this generation than has been possible in the past.

Judge, Puckett and Cabuk (2004) have reported that it is increasingly important for early childhood educators to introduce and use computers in their settings, particularly for those children who do not have access in the home. There is considerable literature that documents the potential of ICTs to create innovative, engaging and substantive learning opportunities for young children.

Researcher sought to find out how students used computers, the skills they were developing, and the links with home and formal learning environments. Researcher undertook this through a survey in which students reported their use of computers.

**Rationale of the study**

Clements (2002) has shown that children working in pairs at the computer engaged more than when working on puzzles on the floor. Yelland (2002) explored the use of computers in the home to develop mathematical ideas and reported that there was considerable potential for computer games to support such learning. Similarly, working at computers has been found to create opportunities for the development of social skills (Lau, 2000). Studies have found that open-ended, child-directed software made a more significant difference in children's developmental gains than did 'drill and practice' software (Haugland, 1997).

**Objective**

1. To study the use of computer by secondary school students.
2. To study the frequency of computer usage by secondary school students.
3. To study the Computer-related skills developed among secondary students.

**Method**

In the present study researcher had collected data from secondary school students of Surat city so survey research method was used.

**Sample**

A survey was implemented in a major area of Surat city. Secondary school students from gujarati medium school in Surat city were population in present study. Through convenient sampling method 200 students (100 girls and 100 boys) were selected from secondary school for data collection.

**Research tool**

Two different scales were used. Some questions sought to identify where and how teens accessed and used computers (and so only a check mark was needed), whereas the frequency of use was documented by Likert scaling, where a 1–3 rating scale was used.

**Data Analysis**

**1. Computer access**

Where and how students were able to access computers and the frequency of that access is discussed here. Some students reported that they had multiple sources of access—a home computer, friends' and at school.

Table1. Access to computers

|       | Home | Friends | School | Cyber cafe | No access |
|-------|------|---------|--------|------------|-----------|
| Girls | 87%  | 32%     | 88%    | 40%        | 12%       |
| Boys  | 90%  | 53%     | 79%    | 45%        | 5%        |

The data in Table 1 suggests that the majority of the respondents access computers in their schools and own homes but can also access them elsewhere. Only a limited number reported having no access to computers.

**2. Frequency of access to computers**

In seeking to understand how students access and use computers and to identify how frequently computers were used in for various functions—playing/recreation, educational purposes and creative purposes one survey question was asked to students. Using a Likert scale where 0=never; 1=sometimes; 2=frequent; 3=regular

Table 2 Use of computers

|  | For internet | For play | For education | For creative work |
|---|---|---|---|---|
| Girls | 76% | 68% | 79% | 60% |
| Boys | 78% | 89% | 76% | 65% |

The data in Table 2 suggests that the majority of the respondents access computers for education, internet and play. Compare to girls Boys use computer more for play. Less number reported having access for creative work.

Table 3 Frequency of access of computer

| Use of computer | Never | | Sometimes | | Frequent | | Regular | |
|---|---|---|---|---|---|---|---|---|
|  | Girls | Boys | Girls | Boys | Girls | Boys | Girls | Boys |
| For play | 32 % | 11% | 44% | 34% | 16% | 41% | 8% | 14% |
| For education | 21% | 24% | 24% | 39% | 40% | 22% | 15% | 15% |
| For creative work | 40% | 35% | 41% | 39% | 13% | 16% | 6% | 10% |
| For internet | 24% | 22% | 43% | 33% | 21% | 26% | 12% | 19% |

The data in Table 3 suggests that girls and boys both are using computer sometimes for creative work and internet. Boys are using computer frequently for play and girls are using computer frequently for education. Only a limited number reported having access to computers regularly.

**3. Activities undertaken when using computers**

By asking students to fill in a check-box, we sought to ascertain the types of activities students engage in when using computers.

Table 4 Activities undertaken while using the computer

| Activities undertaken while using computer | Girls | Boys |
|---|---|---|
| Games | 68% | 89% |
| Drawing | 43% | 33% |
| Subject Software | 49% | 26% |
| Pre-writing activities | 22% | 17% |
| Surfing | 26% | 48% |

The data in table 4 suggests that students are accessing computers in a variety of ways that develop a range of computer skills, knowledge and dispositions. It indicates that there is considerable use of computer games by boys and girls. Many boys use computer for surfing and girls for drawing and subject software.

**4. Computer skills of students**

Having identified how students are accessing computers and for what purposes, researcher sought to identify the skills they were developing as a consequence of their interactions with computers.

Table 5. Computer-related skills developed by students

| Skills | girls | boys |
|---|---|---|
| Type letters | 64% | 59% |
| Retrieve files | 77% | 80% |
| Use of browser | 63% | 70% |
| Use drawing tools | 71% | 74% |
| CD/DVD write | 78% | 81% |
| Use the tool bar | 82% | 86% |
| Print documents/files/screen | 86% | 88% |
| PPT | 25% | 21% |
| Excel | 12% | 18% |

This data of table 5 suggests that students have developed a high number of skills through their interactions with the computer. Many students reported that they can type letters, retrieve files, use browser, use drawing tool, write CD/DVD and print documents very easily but they are not skillful for making of PPT and using of excel.

 **Conclusion**

The data presented in this paper indicates that students have considerable access to computers in school and at home. They access computer for many purposes such as play, internet, and drawing. We should motivate students to use computer for learning and for acquiring knowledge. Students have ability to access computer well but they do have ability to make PPT and use of EXCEL. There is need to focus on development of computer skill among student which are essential for higher study.